\documentclass[floatfix,pra,twocolumn,aps,superscriptaddress,showpacs]{revtex4-1}
\usepackage{hyperref}
\hypersetup{
	colorlinks=true,
	citecolor=blue,
	linkcolor=blue,
	urlcolor=blue}
\usepackage{graphicx}
\usepackage[normalem]{ulem}
\usepackage{mathrsfs,amsmath}
\usepackage{amsfonts}
\usepackage{amssymb}
\usepackage{epsfig}
\usepackage{subfigure}
\usepackage{mathtools}
\usepackage{braket}
\usepackage[usenames,dvipsnames]{color}
\usepackage{setspace}
\usepackage{bm}
\usepackage{times}
\usepackage{ulem}
\usepackage{dsfont}
\usepackage{xcolor}
\usepackage{adjustbox}

\begin{document}
\title[]{Collective excitations and universal coarsening dynamics of a 
spin-orbit-coupled spin-1 Bose-Einstein condensate}
\author{Rajat}\email{rajat.19phz0009@iitrpr.ac.in}
\affiliation{Department of Physics, Indian Institute of Technology Ropar, Rupnagar 140001, Punjab, India}
\author{Paramjeet Banger}\email{paramjeet.banger@acads.iiserpune.ac.in}
\affiliation{Department of Physics, Indian Institute of Technology Ropar, Rupnagar 140001, Punjab, India}
\affiliation{Department of Physics, Indian Institute of Science Education and Research Pune, Pune 411008, India}
\author{Sandeep Gautam}\email{sandeep@iitrpr.ac.in}
\affiliation{Department of Physics, Indian Institute of Technology Ropar, Rupnagar 140001, Punjab, India}

\begin{abstract}
We study the collective excitation spectrum of a Raman-induced spin-orbit-coupled
spin-1 Bose-Einstein condensate confined in a quasi-one-dimensional harmonic trap 
while varying either the Raman coupling or quadratic Zeeman field strength by using the 
Bogoliubov approach. A few low-lying modes, which can be used to delineate the phase boundaries,
are identified by exciting them with suitable perturbations. 
We also investigate the coarsening dynamics of a homogeneous quasi-two-dimensional 
spin-orbit-coupled spin-1 condensate by quenching from the zero-momentum into the 
plane wave phase through a sudden change in Raman coupling or quadratic Zeeman field strength.
We demonstrate that the correlation function of the order parameter displays dynamic
scaling during the late-time dynamics, allowing us to determine the dynamic critical
exponent.

\end{abstract}
\maketitle
\section{Introduction}
One of the remarkable accomplishments in ultracold quantum gases is the creation of artificial gauge potentials, which
allows neutral ultracold atoms to mimic the behavior of charged particles in external electric or magnetic
fields~\cite{lin2011synthetic, Goldman_2014}. Synthetic gauge fields can be designed to interact with internal degrees of
freedom, like atomic spin. This process involves coupling each atom's spin with its center-of-mass motion, commonly
known as spin-orbit (SO) coupling~\cite{galitski2013spin, Goldman_2014}. Experimental realizations of SO
coupling in ultracold neutral atomic gases~\cite{Lin2011, PhysRevLett.109.115301, Campbell2016, luo2016} has provided a novel route to
study exotic quantum phases and nonlinear dynamics~\cite{Zhai_2015}.

In a Raman-induced SO-coupled Bose-Einstein condensate (BEC), various ground state phases can emerge, including
the supersolid stripe (ST) phase, the plane wave (PW) phase, and the zero momentum (ZM)
phase~\cite{PhysRevLett.107.150403, PhysRevLett.108.225301, 10.21468/SciPostPhys.11.5.092}. The ST phase has been
identified as having supersolid properties as it spontaneously breaks both the gauge and continuous translational
symmetry~\cite{li2017stripe, PhysRevLett.124.053605, PhysRevLett.130.156001, chisholm2024probing}. The PW phase breaks ${\mathbb Z}_{2}$
symmetry and features non-zero magnetization~\cite{PhysRevLett.108.225301, PhysRevLett.117.125301, PhysRevA.93.033648}.
The PW and the ST phases correspond to condensation in a single and a pair of plane-wave states, respectively, while in
the ZM phase, BEC occurs in the zero-momentum state~\cite{ PhysRevLett.108.225301}. Since the achievement of BEC, investigating 
collective excitations has become an essential tool for gaining insight into the macroscopic quantum phenomena that
govern these systems~\cite{pethick_smith_2008,pitaevskii2016bose}. In the SO-coupled pseudo-spinor BECs, the excitation
spectrum can demarcate the phase boundaries in both
uniform ~\cite{PhysRevA.86.063621, PhysRevLett.110.235302, PhysRevA.90.063624, PhysRevLett.114.105301} and harmonically trapped
systems~\cite{PhysRevLett.109.115301, chisholm2024probing, PhysRevA.95.033616, PhysRevLett.127.115301, PhysRevLett.130.156001, PhysRevA.109.033319, PhysRevA.109.053307}.
It has now been established that as the Raman coupling strength decreases, the roton gap in the
dispersion of the PW phase also decreases and eventually closes at the boundary between the PW and
ST phases~\cite{PhysRevA.90.063624, PhysRevLett.114.105301, Zheng_2013, PhysRevA.95.033616}. 

The excitation spectrum of the homogeneous SO-coupled spin-1 BECs, which can feature a double roton-maxon structure, has
also been used to discern the phase boundaries between different phases~\cite{PhysRevA.93.033648, PhysRevA.93.023615, Chen_2022}. 
Additionally, it has been observed that at small values of Raman coupling $(\Omega)$ and quadratic Zeeman field
strength $(\epsilon)$, a direct transition from the ZM to the ST phase can occur characterized by symmetric double
rotons~\cite{PhysRevA.93.033648, PhysRevA.93.023615}, which are unique to spin-1 BECs (see also reference ~\cite{PhysRevA.111.023311}). 
As the transition point approaches, the double rotons soften, indicating the system's tendency toward crystallization.
The double roton gap has been used to infer the temperature-induced shift of the ST-ZM phase boundary in a homogeneous SO-coupled 
spin-1 BEC ~\cite{ritu2024thermal}.
The collective excitations of a trapped quasi-one-dimensional (quasi-1D) spin-1 BEC with spin coupling the linear
momentum but crucially, without Raman coupling and quadratic Zeeman field strength have been studied at zero and finite
temperatures~\cite{PhysRevA.106.013304}. Due to the absence of Raman coupling and quadratic Zeeman field strength, typical
features of Raman-induced SO-coupled, like supersolid phase, roton-maxon structure, etc., do not emerge in such a system.
Besides the theoretical studies on the homogeneous SO-coupled spin-1 BECs, the collective
excitation spectrum of a trapped Raman-induced SO-coupled spin-1 BEC~\cite{Campbell2016, luo2016} remains uninvestigated.

Collective modes play an important role in the sudden quench dynamics \cite{PhysRevA.95.023616, huh2024universality}, where
the excitation spectra of the post-quench initial states highlight dynamically unstable modes~\cite{PhysRevA.95.023616}.
It is essential to account for fluctuations beyond the mean-field order to encourage the growth of unstable modes triggered
by the sudden change. The truncated Wigner prescription provides an effective way to introduce fluctuations into the initial
state~\cite{00018730802564254}. When a system undergoes a quench from a disordered to an ordered phase, the growth of order
occurs through the formation of phase domains~\cite{doi:10.1080/00018739400101505}. 
The size of these domains is generally governed by a characteristic length scale, $L(t)$, which increases as the system evolves.
Once this length scale surpasses certain microscopic thresholds, the 
phase ordering process often becomes universal, following a power-law growth
$L(t )\sim t^\beta$, where $\beta$ is dynamic critical exponent. Universal coarsening dynamics, which describe this
growth, have been theoretically 
studied in binary~\cite{ PhysRevLett.113.095702, PhysRevA.101.023608, PhysRevResearch.5.043042} and 
spin-1 spinor BECs~\cite{PhysRevLett.98.160404, *PhysRevLett.99.120407,*PhysRevD.81.025017,*Uhlmann_2010, PhysRevB.76.104519, PhysRevA.88.013630, PhysRevA.91.053609, PhysRevLett.116.025301, PhysRevA.94.023608, PhysRevLett.119.255301, PhysRevA.96.013602, PhysRevA.98.063618,  PhysRevA.100.033603, PhysRevA.95.023616, PhysRevA.99.033611, PhysRevLett.122.173001}, particularly in quasi-1D~\cite{PhysRevA.99.033611,  PhysRevLett.120.073002, PhysRevLett.122.173001, pietraszewicz2021multifaceted} and quasi-two-dimensional (quasi-2D)
configurations for quenching into easy-axis or easy-plane ferromagnetic~\cite{PhysRevA.95.023616, PhysRevLett.116.025301, PhysRevA.94.023608, PhysRevA.100.033603, PhysRevA.99.033611, 10.21468/SciPostPhys.7.3.029} and isotropic phases~\cite{PhysRevLett.119.255301}.
Experimental studies have also confirmed the universal relaxation dynamics in spinor~\cite{prufer2018observation, huh2024universality, PhysRevLett.131.183402} and isolated Bose gases~\cite{erne2018universal, glidden2021bidirectional, PhysRevA.106.023314}. 
However, the universal coarsening dynamics of an SO-coupled spinor BEC with a qualitatively distinct phase diagram have
not been studied.

In the first part of this study, we theoretically investigate the collective excitations of a harmonically trapped quasi-1D
Raman-induced SO-coupled spin-1 BEC as a function of Raman coupling and quadratic Zeeman field strengths. In the second part, we study the 
ensuing dynamics in a homogenous quasi-2D SO-coupled spin-1 BEC in the (non-magnetized) ZM phase after a sudden quench of Raman
coupling or quadratic Zeeman field strength to a value corresponding to which the ground state phase is the (magnetized) PW phase.
We demonstrate that over the extended timescales following the quench, a universal scaling regime emerges, wherein the
order-parameter autocorrelation functions at different times collapse onto a universal scaling function (independent of time) when
the spatial variable is rescaled by a characteristic length $L(t)$, which shows a power law increase with time.   

The paper is organized as follows. In Sec.~\ref{Sec-II}, we introduce the mean-field model to study a
Raman-induced SO-coupled spin-1 BEC. In Sec.~\ref{Sec-IIA}, we present the Bogoliubov-de Gennes (BdG) equations 
for the system. In Sec.~\ref{Sec-IIB}, we discuss the methodology to excite a few low-lying excitations
via suitable perturbations to the Hamiltonian. Sec.~\ref{Sec-III} discusses the phase 
diagram and the collective excitation spectrum of a harmonically confined quasi-1D SO-coupled BEC. 
We then turn our attention to the universal coarsening dynamics of a homogeneous quasi-2D SO-coupled spin-1 BEC in
Sec.~\ref{Sec-IV}, focusing on the sudden quench of Raman coupling, which triggers phase transition from the ZM phase
to the PW phase. Finally, in Sec.~\ref{Sec-V}, we conclude with a summary of the key findings of this study.

\section{Model}
\label{Sec-II}
The ground state properties of an SO-coupled spin-1 BEC can be analyzed by minimizing the energy 
functional~\cite{Campbell2016, luo2016, PhysRevA.89.023630, PhysRevLett.117.125301}
\begin{equation}
E[\Phi]=\int d {\bf r}\left\{\Phi^{\dagger} H_0 \Phi+\frac{c_0}{2}\left(\Phi^{\dagger} \Phi\right)^2+\frac{c_2}{2}\left(\Phi^{\dagger} \boldsymbol{{\rm S}} \Phi\right)^2\right\},
\label{en_fxnl}
\end{equation} 
where $m$ is the atomic mass,
\begin{align}
\label{sph}
H_0=&\frac{\hbar^2}{2 m}\left(-\iota \frac{\partial}{\partial x}+2 k_R {\rm S}_z\right)^2-\frac{\hbar^2}{2 m} \left(\frac{\partial^2}{\partial y^2}+ \frac{\partial^2}{\partial z^2}\right)+V({\bf r})\nonumber\\&
+\Omega {\rm S}_x+\epsilon {\rm S}_z^2,
\end{align}
is the single-particle Hamiltonian, $\Phi$ is the three-component condensate wave function with $\int |\Phi|^2d{\bf r}$ equal to the number
of atoms $N$, $V({\bf r})$ is the harmonic oscillator potential, $\boldsymbol{{\rm S}} = ({\rm S}_x, {\rm S}_y, {\rm S}_z)$ is a vector of 
spin-1 matrices, $c_0= 4 \pi \hbar^2(a_0 +2 a_2)/3m$ and $c_2= 4 \pi\hbar^2 (a_2 - a_0)/3m$ stand for spin-independent and spin-dependent 
interactions, written in terms of the $s$-wave scattering lengths $a_0$ and $a_2$ of binary collisions with total spin equal to $0$ and 
$2$, respectively. In Eq.~(\ref{sph}), $\Omega$ and $k_R$ are the Raman and spin-orbit coupling strengths, respectively, and $\epsilon$ is 
the quadratic Zeeman field strength \cite{Campbell2016}. Depending on the sign of $c_2$, a spin-1 BEC can have
ferromagnetic ($c_2<0$) or antiferromagnetic ($c_2>0$) interactions, and in this manuscript, we consider a spin-1 BEC with antiferromagnetic interactions. To study the dynamics of the system, it is convenient 
to introduce
the Lagrangian
\begin{equation}
L = \int d{\bf r}\frac{i\hbar}{2} \left(\Phi^{\dagger}\frac{\partial \Phi}{\partial t} - \Phi\frac{\partial \Phi^\dagger}{\partial t}\right) - E[\Phi].
\end{equation}

In the first part of this manuscript, we consider a quasi-1D SO-coupled spin-1 BEC 
of $^{23}$Na in a highly anisotropic axisymmetric harmonic trap with significantly stronger confinement along
the radial than the axial direction. 
The action principle leads to the coupled Gross-Pitaevskii equations (GPEs), which in the dimensionless form are
~\cite{PhysRevA.65.043614, Campbell2016, luo2016}
\begin{subequations}\label{gpe}
\begin{align}
\iota\frac{\partial \phi_{\pm 1}}{\partial t} =&\left[ -\frac{1}{2}\partial_x^2 + 2 k_{R}^2 \mp \iota 2 k_{R} \partial_x + V(x)  +\epsilon + c_{0}n \right]\phi_{\pm1}\nonumber\\
& + {c_{2}(n_{\pm1}+n_{0}-n_{\mp1})}\phi_{\pm1}+c_{2}\phi_0^2\phi_{\mp1}^* + \frac{\Omega}{\sqrt{2}} \phi_0 ,\\
\iota\frac{\partial \phi_0}{\partial t}= &\left[-\frac{1}{2}\partial_x^2 + V(x) + c_{0}n+c_{2}(n_{+1}+n_{-1})\right]\phi_0\nonumber\\
&+2c_{2}\phi_{+1}\phi_0^*\phi_{-1}+\frac{\Omega}{\sqrt{2}}\left(\phi_{+1}+\phi_{-1} \right),
\end{align}
\end{subequations}
where $V(x) = x^2/2$ is the harmonic trapping potential along the axial $x$-direction, $n_j = |\phi_j(x)|^2$ and $n=\sum_j n_j$ are
the component and total densities, respectively, with $\int n(x) dx = 1$. In Eqs.~(\ref{gpe}a) and (\ref{gpe}b)
time, length, and energy are in the units of $\omega_{x}^{-1}$, $l_0 = \sqrt{\hbar/m\omega_x}$, and  $\hbar \omega_{x}$, respectively, 
where $\omega_x$ is the trap frequency along the weakly confined axial direction. The dimensionless interaction strengths are
$c_{0} =2 (a_0 +2 a_2)\kappa/(3 l_0)$ and $ c_{2} =2 (a_2 - a_0) \kappa/(3 l_0)$ with $\kappa$
as the ratio of the trap frequencies along the radial direction to the axial direction. For the stationary solutions, 
substituting $\phi_{j}(x,t) = \phi_j(x)e^{-\iota\mu t}$, where $\mu$ denotes the chemical potential, in Eqs.~(\ref{gpe}a)-(\ref{gpe}b) yields the time-independent version of
the GPEs.
To investigate the collective excitation spectrum of the quasi-1D SO-coupled BEC, first, we use the
Bogoliubov-de Gennes (BdG) equations and then validate our results by exciting a few collective modes with suitable perturbations to the Hamiltonian. 
\subsection{Bogoliubov-de Gennes (BdG) equations}
\label{Sec-IIA}
We employ the Bogoliubov approach to investigate the collective excitation spectrum.
Here, one incorporates the fluctuations to the ground state leading to the perturbed
order parameter
\begin{equation}
    \phi_j(x,t) = e^{-\iota\mu t} [\phi_j(x) +\delta {\phi}_j(x,t)],
\end{equation}
where $\phi_j(x)$ is the $j$th component's ground-state wavefunction, $\delta{\phi}_j(x,t) = u_j^{\lambda}(x) e^{-\iota\omega_{\lambda} t}-v_j^{\lambda*}(x) e^{\iota\omega_{\lambda} t}$ with $u_j^{\lambda}(x)$ and 
$v_j^{\lambda}(x)$ denoting the Bogoliubov amplitudes and $\omega_{\lambda}$ the excitation frequency. The GPEs (\ref{gpe}a) and (\ref{gpe}b) are 
linearized to obtain BdG equations, 
\begin{equation}
\begin{pmatrix}
\mathcal{A} & -\mathcal{B}\\
\mathcal{B}^* & -\mathcal{A}^*
\end{pmatrix} 
\begin{pmatrix}
{\mathbf u}^{\lambda} \\ {\mathbf v}^{\lambda} 
\end{pmatrix} 
=\omega_{\lambda}
\begin{pmatrix}
{\mathbf u}^{\lambda} \\{\mathbf v}^{\lambda} 
\end{pmatrix},
\label{bdg}
\end{equation}
where ${\mathbf u}^{\lambda} = \left(u_{+1}^{\lambda},u_{0}^{\lambda},u_{-1}^{\lambda}\right)^{T}$,
${\mathbf v}^{\lambda} = \left(v_{+1}^{\lambda},v_{0}^{\lambda},v_{-1}^{\lambda}\right)^{T}$, $\cal A$ and $\cal B$ are $3\times3$ matrices, and $^*$ denotes the complex conjuate.
The elements of the $\cal A$ and $\cal B$ are defined as follows: 
\begin{align*}
{\cal A}_{11} =&\left(-\frac{1}{2} \partial_x^2-\mu +\epsilon + 2 k_{R}^2+V(x)+c_{0}n\right)\nonumber\\
               &+ c_{0} n_{+1}+{c_{2}(2 n_{+1}+n_{0}-n_{-1})}-\iota 2 k_{R} \partial_x,\nonumber\\
{\cal A}_{12} =& (c_{0}+c_{2})(\phi_0^*\phi_{+1})+2 c_2(\phi_{-1}^*\phi_{0})+\frac{\Omega}{\sqrt{2}},\nonumber\\ 
{\cal A}_{13}=&{(c_{0}-c_{2})(\phi_{-1}^*\phi_{+1})},\nonumber\\
{\cal A}_{22} =&\left(-\frac{1}{2} \partial_x^2-\mu+V(x)+c_{0}n\right)\nonumber\\
               &+c_{0} n_{0}+{c_{2}(n_{+1}+n_{-1})},\nonumber\\
{\cal A}_{23} =& (c_{0}+c_{2})(\phi_{0}\phi_{-1}^*)+2c_2(\phi_{+1}\phi_{0}^*)+\frac{\Omega}{\sqrt{2}},\nonumber\\
{\cal A}_{33} =&\left(-\frac{1}{2} \partial_x^2-\mu +\epsilon + 2 k_{R}^2+V(x)+c_{0}n\right)\nonumber\\
               &+ c_{0} n_{-1}+{c_{2}(2 n_{-1}+n_{0}-n_{+1})}+\iota 2 k_{R} \partial_x,\nonumber\\               
{\cal A}_{21} =& {\cal A}_{12}^*,\quad {\cal A}_{31} = {\cal A}_{13}^*,\quad {\cal A}_{32} = {\cal A}_{23}^*,\\
{\cal B}_{11}=&{(c_{0}+c_{2})\phi_{+1}^2}, \quad{\cal B}_{12}={(c_{0}+c_{2})\phi_0\phi_{+1}}+\frac{\Omega}{\sqrt{2}},\nonumber\\
{\cal B}_{13}=&{(c_{0}-c_{2})\phi_{-1}\phi_{+1}+c_{2}\phi_0^2},\nonumber\\
{\cal B}_{22}=&{c_{0}\phi_{0}^2+2 c_2 \phi_{+1} \phi_{-1}},\nonumber\\
{\cal B}_{23}=&{(c_{0}-c_{2})\phi_{+1}\phi_{-1}+c_{2}\phi_0^2}+\frac{\Omega}{\sqrt{2}},\nonumber\\
{\cal B}_{33}=&(c_{0}+c_{2})\phi_{-1}^{2},\nonumber\\
{\cal B}_{21}=& {\cal B}_{12},\quad {\cal B}_{31} = {\cal B}_{13},\quad {\cal B}_{32} = {\cal B}_{23}.\\
\end{align*}
We solve one-dimensional GPEs (\ref{gpe}a) and (\ref{gpe}b) numerically using a split time-step Fourier pseudospectral method
and calculate the ground-state solution using imaginary-time propagation \cite{kaur2021fortress,*banger2022fortress,*banger2021semi}.
We consider a spatial step size $\Delta x = 0.01$ and an imaginary-time step of $10^{-5}$ in the imaginary-time propagation. 
We solve the BdG equations for a harmonically confined BEC by expanding the quasiparticle amplitudes 
in terms of the eigenfunctions of the one-dimensional harmonic oscillator, which leads to a generalized matrix eigenvalue
problem for the expansion coefficients~\cite{roy-2020, PhysRevA.111.023311}. We then solve this eigenvalue problem using conventional matrix
diagonalization subroutines~\cite{lapack, doi:10.1137/1.9780898719628} to obtain the eigenenergies ($\omega_{\lambda}$) and
quasiparticle amplitudes ($u^{\lambda}$ and $v^{\lambda}$). In this work, we consider a truncated basis set of $180$ low-lying
harmonic oscillator eigenstates to expand quasiparticle amplitudes. We have confirmed that increasing the basis size does not
affect the results for low-lying collective excitations of the quasi-1D system considered in this work.
\subsection{Time-dependent Gross-Pitaevskii
equation with perturbations}
\label{Sec-IIB}
A few low-lying collective excitations of a spinor BEC, like dipole, spin-dipole, breathing, and spin-breathing modes, can
be excited by perturbing the underlying Hamiltonian with a suitable perturbation~\cite{PhysRevLett.77.988, PhysRevA.94.063652}.
These modes are of great interest and can be investigated in experiments.
Conceptually, to excite these modes, one can perturb the trapping potential as follows:
\begin{equation}
V(x) = 
 \begin{cases}\label{mod_pot}
     \frac{1}{2} \left (x  + \delta \times \hat{O} \right)^2\quad\textrm{for dipole modes}\\ 
     \frac{1}{2} \left (x^2 + \delta \times\hat{O} \right)\quad\textrm{for breathing modes}     
 \end{cases}.
\end{equation}
where $\hat O$ is an observable and $\delta$ is a small real number. Depending on the mode to be excited, the observable $\hat O$
is of form
\begin{equation}
 \hat O = 
\begin{cases} 
x\quad \textrm{for the dipole mode}\\
x {\rm S}_z\quad \textrm{for the spin-dipole mode} \\
x^2 \quad \textrm{for the breathing mode}\\
x^2 {\rm S}_z \quad \textrm{for the spin-breathing mode} 
\end{cases}.
\end{equation} 
We first calculate the ground-state solution of the quasi-1D SO-coupled BEC under spin-independent harmonic 
confinement $V(x) = x^2/2$ and consider this as the solution of GPEs (\ref{gpe}a) and (\ref{gpe}b) at $t=0$. For $t>0$, we suddenly switch to potential 
$V(x)$ in Eq.~(\ref{mod_pot}) and examine $\langle \hat O\rangle = \int \Phi^\dagger(x,t) \hat O \Phi(x,t) dx$ as a function of 
time and extract the collective excitation frequency from its Fourier transform. 
\section{Ground state phases and Collective Excitations}
\label{Sec-III}

The general ground state solution of a quasi-1D Raman-induced SO-coupled spin-1 BEC is of the form~\cite{PhysRevLett.117.125301}
\begin{equation}
\Phi(x) = \sqrt{n_0(x)}\sum_{l\in \mathbb Z} C_l \zeta_l e^{\iota lk x},
\label{gs_wave_fxn}
\end{equation}
where $n_0(x)$ is the total density in the absence of SO-coupling, $k$ is the condensate's momentum, $\zeta_l$ are three-component 
normalized spinor,
and complex coefficients $C_l$ satisfy $\sum_{l\in \mathbb Z}|C_l|^2 = 1$.
In this section, we consider an SO-coupled spin-1 BEC of $^{23}$Na atoms with $a_{0}= 48.91 a_{B}$ and
$a_{2}=54.54 a_{B}$~\cite{PhysRevA.83.042704}, where $a_{B}$ is the Bohr radius, and SO-coupling strength $k_R = 3$.
The BEC is confined in a quasi-1D trap with the trapping frequencies $\omega_x=2\pi\times5$ Hz and $\kappa = 20$. The (dimensionless) 
interaction strengths translate to $c_{0}=59.44$ and $c_{2}=2.12$ for the number of atoms $N=5000$.
\begin{figure}[!htbp]
    \centering
        \includegraphics[width=\columnwidth]{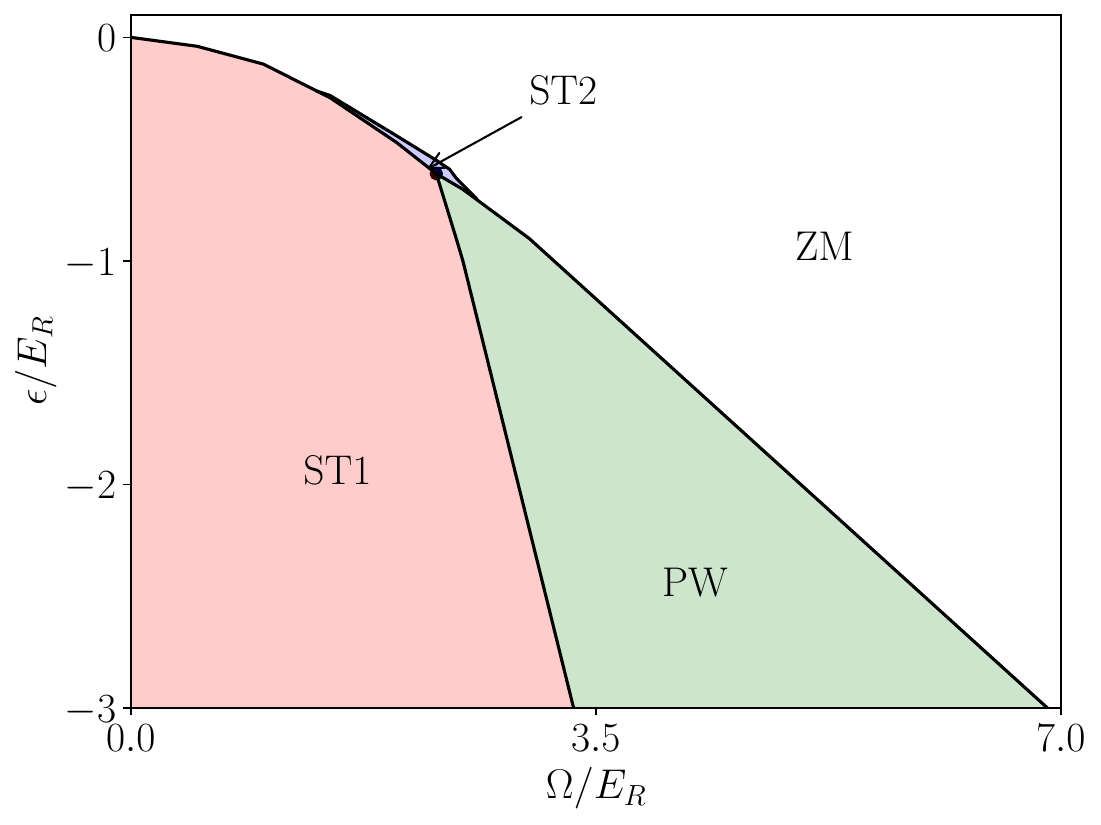}
    \caption{Phase diagram of a harmonically-trapped SO-coupled spin-1 BEC with $c_0= 59.44$, $c_2= 2.12$ and $k_R=3$. 
    A filled circle indicates the tricritical point.}
    \label{gs_phase}
\end{figure}
In Fig.~\ref{gs_phase}, we show the ground-state phase diagram with two distinct supersolid ST phases, denoted
by ST1 and ST2, the PW phase and the ZM phase. The supersolid ST1 phase features occupation of only odd$-l$ states
in Eq.~(\ref{gs_wave_fxn}), while the supersolid ST2 phase, occupying a minuscule region of the parameter space, 
features both odd and even$-l$ states~\cite{PhysRevLett.117.125301}.
Consequently, in the ST2 phase, longitudinal magnetization density $n_{+1}(x) -n_{-1}(x)$ 
oscillates with a period of $2 \pi/k$, whereas in the ST1, it oscillates with a period of $\pi/k$. However, the overall density in both ST1 and ST2 phases oscillates with a period of $\pi/k$. The phase diagram is qualitatively similar to the phase diagram for the homogeneous SO-coupled spin-1 BEC
with antiferromagnetic interactions in Ref.~\cite{PhysRevA.93.033648}; however, the trap-induced inhomogeneity and different interaction strengths 
lead to the shift in the phase boundaries.

\begin{figure}[!htbp]
    \centering
        \includegraphics[width=\columnwidth]{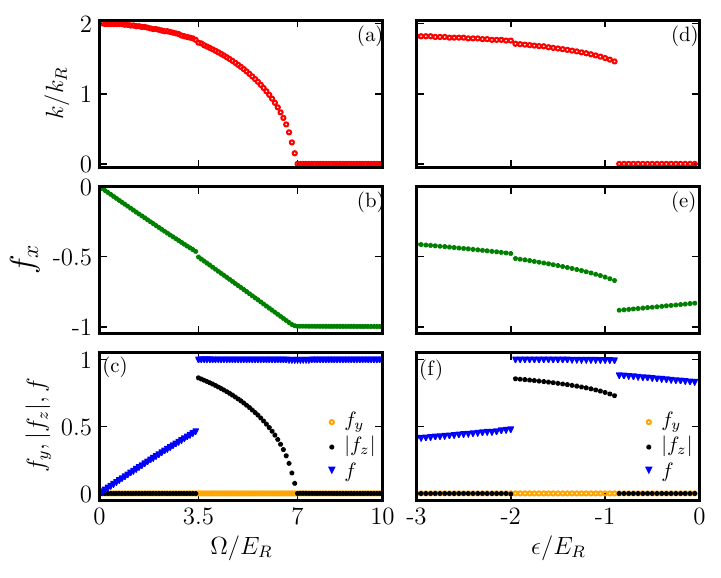}
    \caption{The condensate's momentum and spin-expectation per particle in the ground-state phase for $c_0 = 59.44$, $c_2 = 2.12$, and
     $k_R = 3$: (a)-(c) as a function of $\Omega$ for a fixed $\epsilon = -3E_R$ and (d)-(f)as a function of $\epsilon$ for
     a fixed $\Omega = 3E_R$. (a) and (d) momentum $k$, (b) and (e) $f_x = \int F_x dx$, and (c) and (f) $f_y = \int F_y dy,
     |f_z| = |\int F_z dz|$ and $f$ as a function of $\Omega$. In (a)-(c), $k$, $f_x$, $|f_z|$, and $f$ are discontinuous at
     the ST1-PW phase boundary $\Omega_{\rm c_1}$ but are continuous at the PW-ZM phase boundary $\Omega_{\rm c_2}$, which
     illustrates the first-order and second-order natures of these two transitions, respectively. In (d)-(f), across both the critical
     points, the quantities change discontinuously.}
    \label{phase_diagram} 
\end{figure}

We now first consider quadratic Zeeman field strength $\epsilon = -3 E_{\rm R}$, where $E_R = k_R^2/2$ is the recoil energy, and vary Raman coupling 
strength $\Omega$. 
As $\Omega$ is progressively increased, the system first undergoes a phase transition from the ST1 to the PW phase above 
a critical coupling $\Omega_{\rm c_1} \approx 3.4 E_{\rm R}$ and then a phase transition from the PW to the ZM phase above a critical coupling $
\Omega_{\rm c_2} \approx 6.9 E_{\rm R}$. In Figs.~\ref{phase_diagram}(a)-(c), we show momentum $k$, components of spin-expectation 
per particle 
$f_{\nu} = \int F_{\nu} dx$ with $\nu=x,y,z$, and $f = \sqrt{f_x^2+f_y^2+f_z^2}$ as a function of Raman coupling strength $\Omega$, where 
$F_{\nu} = \Phi(x)^\dagger{\rm S}_\nu\Phi(x)$ is the $\nu$ component of the spin-density vector with order parameter
$\Phi = (\phi_{+1},\phi_0,\phi_{-1})^{T}$.

Across the ST1-PW transition point, $k$, $f_x$, $|f_z|$, and $f$ change discontinuously. The condensate's momentum $k$ is non-zero
in the ST1 and PW phases only [see Fig.~\ref{phase_diagram}(a)]; $|f_x|$ increases linearly with an increase in $\Omega$ in the ST1
and in the PW phase, while it remains constant in the ZM phase [see Fig.~\ref{phase_diagram}(b)]; $|f_z|$ is non-zero only in the
PW phase and $f_y$ is zero across all the three phases [see Fig.~\ref{phase_diagram}(c)]; and $f$ increases linearly with $\Omega$
in the ST1 phase, jumps to $1$ in the PW phase, and then remains equal to $1$ across the PW and the ZM phases 
[see Fig.~\ref{phase_diagram}(c)]. The variation of condensate's momentum and spin expectation per particle as a function
of $\Omega$ is qualitatively identical to the observations made for a homogeneous SO-coupled pseudospinor 
BEC~\cite{PhysRevLett.108.225301}. If, rather, we fix $\Omega = 3 E_R$, momentum $k$ and spin-expectation per particle as a function of 
$\epsilon$ are shown in  Figs.~\ref{phase_diagram}(d), (e), and (f), respectively. 
In this case, too, the system first transitions from the ST1 to the PW above $\epsilon_{\rm c_1}$ and then from the PW to the
ZM phase above $\epsilon_{\rm c_2}$, albeit with a first-order transition across both transition points [as demonstrated by discontinuities in Figs.~\ref{phase_diagram}(d)-(f)].  

\begin{figure}[!htbp]
\includegraphics[width=\columnwidth]{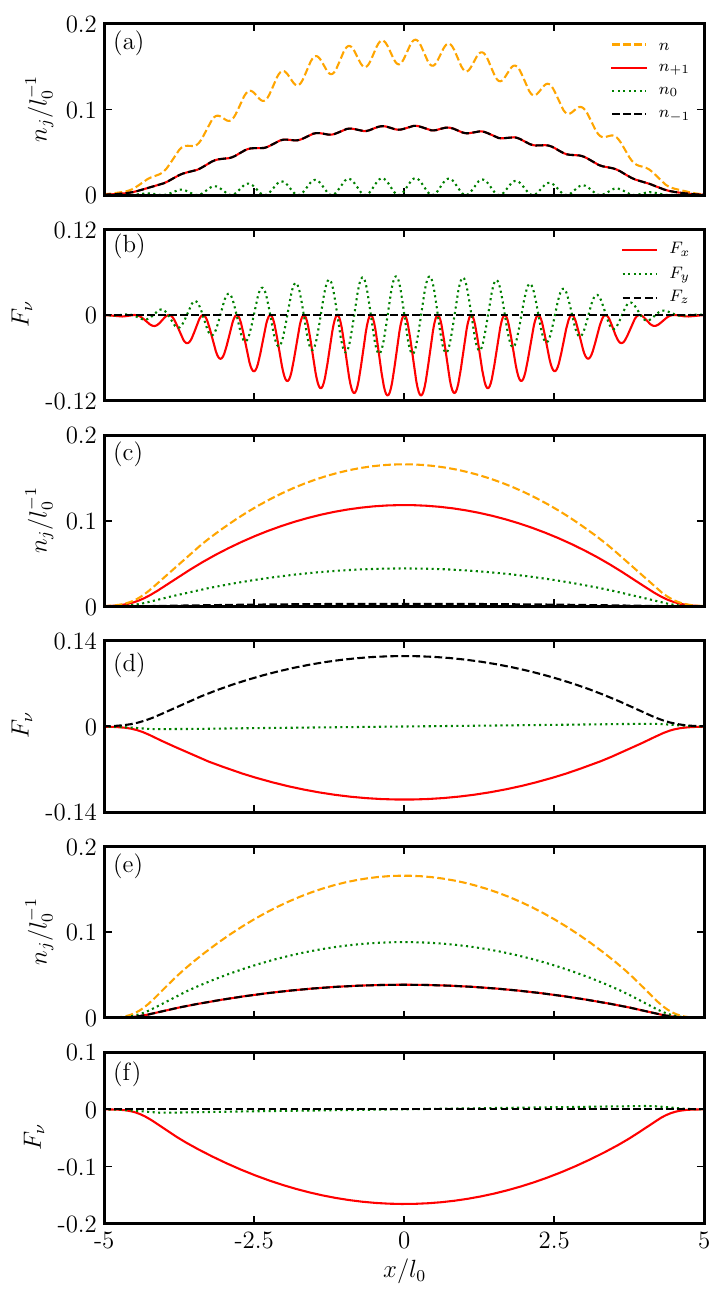}
\caption{Ground-state density and spin-density profiles of the SO-coupled spin-1 BEC with $c_0= 59.44$, $c_2= 2.12$, $k_{\rm R}= 3$, 
and $\epsilon = -3E_R$ in the three phases. (a) and (b) display the density and spin-density profiles in the ST1 phase
for $\Omega = 2.5 E_{\rm R}$, (c) and (d) show the same in the PW phase for $\Omega=5 E_{\rm R}$, and (e) and (f) illustrate
the respective densities in the ZM phase for $\Omega=7.5 E_{\rm R}$.} 
\label{density}
\end{figure}

In Fig.~\ref{density}, we plot the density and the spin-density profiles for the three phases, namely
the ST1 phase at $\Omega=2.5 E_{\rm R}$, the PW phase at $\Omega=5 E_{\rm R}$, and the ZM phase at $\Omega=7.5 E_{\rm R}$ for
a fixed $\epsilon = -3 E_{\rm R} $. 
The component and the total densities oscillate in phase across the spatial extent of the BEC 
for the ST1 phase in Fig.~\ref{density}(a), which is typical for a supersolid.
The densities $n_{+1}(x)$ and $n_{-1}(x)$ overlap resulting in $F_z =0$, and $F_x$ and $F_y$ 
oscillate with the same period as that for densities in this phase [see Fig.~\ref{density}(b)]. 
The PW phase with non-overlapping component densities exhibits non-zero $F_x$ and $F_z$ [cf. Figs.~\ref{density}(c) and \ref{density}(d)]. 
The ZM phase again with overlapping $n_{+1}(x)$ and $n_{-1}(x)$ has $F_z = 0$, whereas $F_x \ne 0$ contributes to yield $f=1$ for this phase [see Figs.~\ref{density}(e) and \ref{density}(f)]. 

\begin{figure}[!htbp]
\centering
\includegraphics[width=\columnwidth]{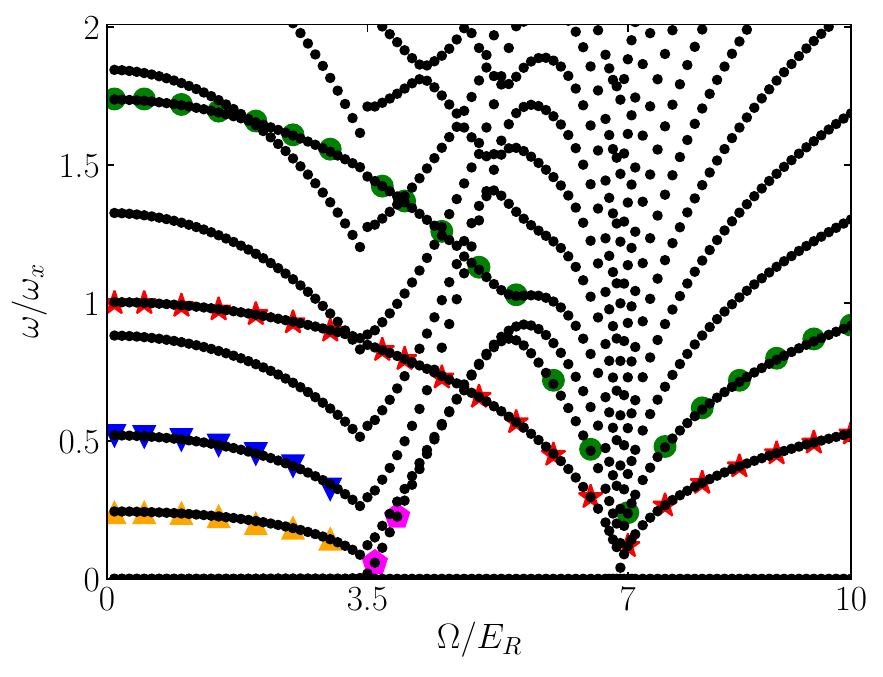}
\caption{Low-lying collective excitations of $^{23}$Na Raman-induced SO-coupled spin-1 BEC with $c_0= 59.44$, $c_2= 2.12$, $k_R=3$, and  
$\epsilon =-3 E_{R}$ as a function of $\Omega$. The density-dipole and density-breathing modes are marked by red asterisks and green-filled circles, 
respectively. The yellow up-triangles and blue down-triangles indicate the spin-dipole and spin-breathing modes, respectively. 
At $\Omega_{\rm c_1}\approx3.4 E_{R}$, there is a phase transition from the ST1 to the PW phase. Several avoided crossings, including the one involving the breathing mode, are evident in the PW phase. The density dipole mode approaches zero at $\Omega_{\rm c_2}\approx6.9 E_{R}$ above which the PW phase transitions to the ZM phase. In the PW phase, the roton excitation vanishing at the PW-ST1 phase boundary is marked by a magenta-coloured filled pentagon.}
\label{varying_omega}
\end{figure}

{\em Collective excitations}: The phase transition in the SO-coupled spin-1 BEC can be driven varying coupling strength $\Omega$ or quadratic Zeeman field strength $\epsilon$~\cite{PhysRevLett.117.125301} . We calculate the excitation spectrum of the quasi-1D BEC as a function of these control parameters
by solving the BdG Eqs.~(\ref{bdg}) as discussed in Sec.~\ref{Sec-IIA} and confirming the nature and the magnitude of a few low-lying excitations 
as discussed in Sec. \ref{Sec-IIB}.
In Fig.~\ref{varying_omega}, we show the excitation spectrum of the quasi-1D SO-coupled BEC as a function of coupling
strength $\Omega$ while the $c_0$, $c_2$, $k_R$, and $\epsilon$ are the same as in Figs.~\ref{phase_diagram}(a)-(c)
With an increase in $\Omega$, the low-lying collective excitations
like spin-dipole, spin-breathing, density-dipole, and density-breathing modes, decrease in the ST1 phase. The spin-dipole and spin-breathing modes (two of the lowest non-zero energy collective excitations) have minimum energies at $\Omega_{\rm{c}_1} \approx 3.4 E_{R}$. The dipole and the breathing modes' excitation frequencies display discontinuities at the ST1-PW phase boundary. Within the PW phase, these two modes continue to decrease with an increase in $\Omega$ and acquire their minimum values at the PW-ZM phase boundary, $\Omega_{\rm{c}_2} \approx 6.9 E_{R}$. In the ZM phase, the two density modes increase with an increase in $\Omega$. The ST1 phase has two zero-energy Goldstone modes corresponding to the breaking of continuous $U(1)$ gauge and translational symmetries. In contrast, the PW and ZM phases have one Goldstone mode due to the breaking of $U(1)$ gauge symmetry.

In an infinite homogeneous SO-coupled BEC, theoretical distinctions between these phases rely on calculating the dynamic structure factor and
sound velocities for spin and density waves~\cite{PhysRevLett.130.156001}. The confinement-induced discrete collective (shape) oscillations were leveraged to probe phase transitions in harmonically trapped SO-coupled pseudospinor BECs; for example, softening of the spin-dipole (stripe compression) 
mode at the supersolid ST-to-PW phase boundary was revealed in a recent experiment by Chisholm {\em et al.} \cite{chisholm2024probing}, similar to softening of the dipole mode at the PW and ZM phases~\cite{PhysRevLett.109.115301,chisholm2024probing}.
The sound velocities for the spin and density waves in a homogeneous SO-coupled spin-1 BEC in Ref.~\cite{Chen_2022} qualitatively mimic the behavior of spin-dipole and
breathing modes in Fig.~\ref{varying_omega}. Pertinently, direct measurements of the sound velocities for spin and density waves in the SO-coupled supersolid phase have not been performed
in experiments.

\begin{figure}[!htbp]
\includegraphics[width=\columnwidth]{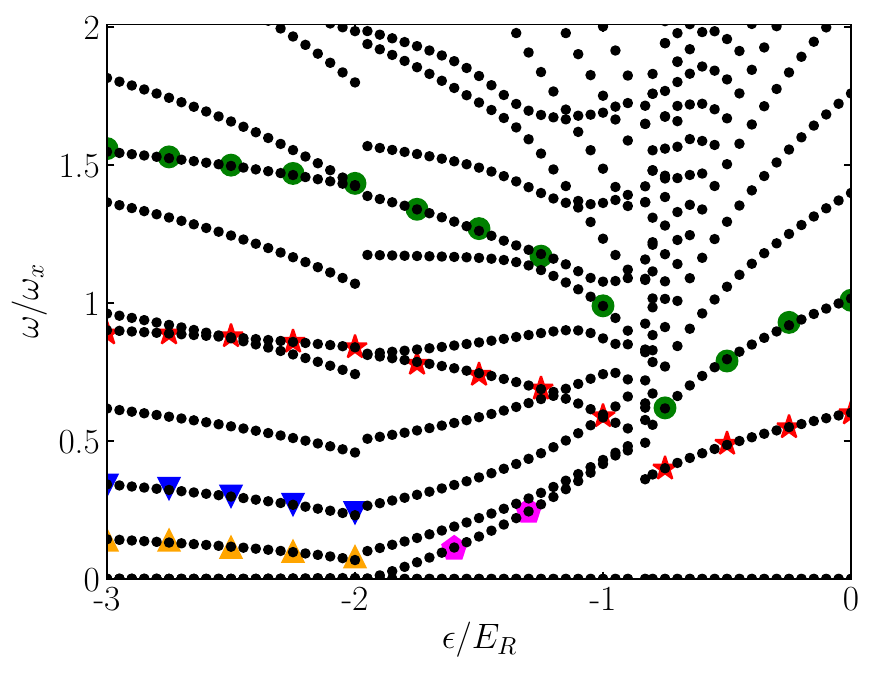}
\caption{
Excitation spectrum of the SO coupled spin-1 BEC with $c_0= 59.44$, $c_2= 2.12$, $k_R=3$, and  
$\Omega = 3 E_{R}$ as a function of quadratic Zeeman field strength $\epsilon$. The density-dipole and density-breathing modes are marked by red asterisks and green-filled circles, 
respectively. The yellow up-triangle and blue down-triangle indicate the spin-dipole and spin-breathing modes, respectively. At $\epsilon_{\rm c_1} \approx -2 E_{R}$, the ST1 phase transitions to the PW phase, and at $\epsilon_{\rm c_2} \approx -0.9 E_{R}$
the PW phase transitions to the ZM phase. The discontinuous jumps in the dipole and the breathing modes across these transition points indicate the first-order nature of these transitions. In the PW phase, the roton excitation vanishing at the PW-ST1 phase boundary is marked by a magenta-coloured filled pentagon.}
\label{varying_zeeman}
\end{figure}

To study the collective excitations as a function of the quadratic Zeeman field strength next, we fix the Raman coupling strength at $\Omega = 3 E_{R}$ and vary the quadratic Zeeman strength 
$\epsilon$. The excitation spectrum as a function of $\epsilon$ is shown in Fig.~\ref{varying_zeeman}.
Similar to the results shown in Fig.~\ref{varying_omega}, within the ST1 phase, the spin-dipole and the spin-breathing modes soften with an
increase in Zeeman strength and have their minimum values at the ST1-PW phase boundary, $\epsilon_{\rm c_1}\approx-2 E_{R}$; the dipole
and breathing modes, which decrease with increasing $\epsilon$ across the ST1 and the PW phases, exhibit discontinuities across the first-order
ST1-PW phase transition point. However, the dipole and breathing modes jump discontinuously across the PW-ZM phase boundary
at $\epsilon_{\rm c_2}\approx-0.9 E_{R}$, in contrast to the spectrum shown in Fig.~\ref{varying_omega}, and then increase with
increasing $\epsilon$ across the ZM phase.
This discontinuity in dipole and breathing modes indicates a first-order transition from the PW phase to the ZM phase. Another distinctive
feature in this case: at $\Omega_{\rm c_2}$ in Fig.~\ref{varying_omega}, dipole mode approaches zero, but there is a finite
gap in Fig.~\ref{varying_zeeman} at $\epsilon_{\rm c_2}$. 

\begin{figure}[!htbp]
\includegraphics[width=\columnwidth]{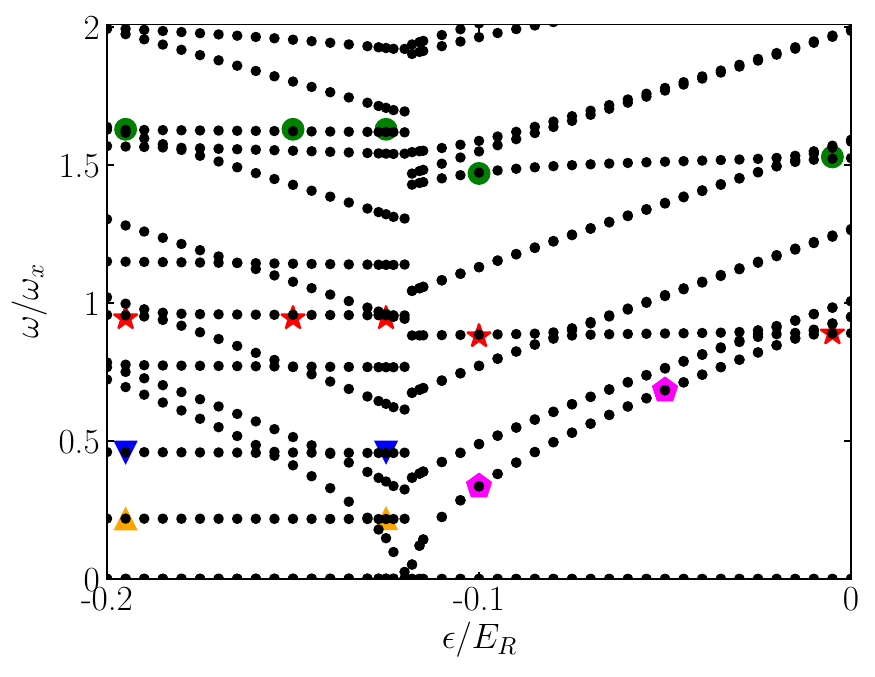}
\caption{Low-lying collective excitations of $^{23}$Na Raman-induced SO coupled spin-1 BEC with $c_0=59.44$, $c_1=2.12$, $\Omega = E_{R}$, and $k_R=3$ as a function of
quadratic Zeeman field strength $\epsilon$.  The density-dipole and density-breathing modes are marked by red stars and green-filled circles, 
respectively. The yellow up-triangles and blue down-triangles indicate the spin-dipole and spin-breathing modes, respectively. 
In the ZM phase, the softening of the lowest-lying double roton modes marked by magenta colored-filled pentagon indicates a phase transition from the ZM phase to the ST1 phase at $\epsilon_{\rm c_1}\approx-0.12 E_{R}$. The dipole and breathing modes vary discontinuously across the first-order phase transition.
}
\label{low_zeeman}
\end{figure}

At lower values of $\Omega$, the phase transition directly from the ST1 to the ZM phase is observed as one varies $\epsilon$~\cite{PhysRevLett.117.125301}. In Fig.~\ref{low_zeeman}, we plot the excitation spectrum of the BEC at a lower $\Omega = E_R$ as a function $\epsilon$. Within the ZM phase, as the quadratic Zeeman strength is decreased,
symmetric double roton modes decrease and vanish at the ZM-ST1 phase boundary, $\epsilon_{\rm{c}_1}\approx-0.12 E_{R}$.
These symmetric double roton modes have been studied for the homogeneous system~\cite{PhysRevA.93.033648, PhysRevA.93.023615, Chen_2022}, but here, their trapped counterparts manifest as discrete excitations.
The dipole, breathing, spin-dipole, and spin-breathing modes remain nearly constant in the ST1 phase, with the two density modes displaying discontinuities at (the first-order) transition point.
 
\section{Coarsening dynamics in a quasi-2D SO-coupled BEC}
\label{Sec-IV}

In this section, we consider the coarsening dynamics~\cite{PhysRevLett.116.025301, PhysRevA.94.023608} in a homogenous 
quasi-2D SO-coupled spin-1 BEC initiated by a sudden quench of the coupling strength, resulting in the transition from the 
ZM phase to the PW phase. In the absence of any confinement, we measure length, time, and energy in units of 
$\zeta_{\rm so} = 1/k_R$, $t_{\rm so} = m / (\hbar k_R^2)$, and $\hbar^2 k_R^2 / m$, respectively, rather than the 
harmonic-trap-based units considered in the previous two sections. 

The ground state wave functions in the PW and ZM phases are of the form  
\begin{equation}\label{PW_ZM_OP} 
\Phi(x,y)=\left[\begin{array}{l}
\phi_{+1}(x,y) \\
\phi_0(x,y) \\
\phi_{-1}(x,y)
\end{array}\right]=\sqrt{n} \mathrm{e}^{\iota k x}\left(\begin{array}{c}
\eta_{+1} e^{\iota \theta_{+1}} \\
\eta_0 \\
\eta_{-1} \mathrm{e}^{\iota \theta_{-1}}
\end{array}\right),
\end{equation} 
where $n=N/A$ is the atom density (number of atoms per unit area), 
and $k$ is the condensate's momentum, which is zero for the ZM and non-zero for the PW phase, 
and $(\eta_{+1},\eta_0,\eta_{-1})$ is a spinor with $\sum_{j=\pm1,0} |\eta_j|^2 = 1$ ~\cite{Chen_2022}. 
The energy per particle  in the PW (or ZM) phase, obtained by substituting $\Phi$ in (\ref{PW_ZM_OP}) in Eq.~(\ref{en_fxnl}), 
is
$$
\begin{aligned}
\frac{E}{N}= & \frac{k^2}{2}+2 k\left(\eta_{+1}^2-\eta_{-1}^2\right)+(\epsilon+2)\left(\eta_{+1}^2+\eta_{-1}^2\right)+\frac{c_0 n}{2}\\
&+\sqrt{2} \Omega \eta_0\left(\eta_{+1} \cos \theta_{+1}+\eta_{-1} \cos \theta_{-1}\right)+\frac{c_2 n}{2}\left(1-2 \eta_{-1}^2\right)^2 \\
& +\frac{c_2 n \eta_0^2}{2}\left[4 \eta_{+1} \eta_{-1} \cos \left(\theta_{+1}+\theta_{-1}\right)+4 \eta_{-1}^2-\eta_0^2\right].
\label{ener_fun}
\end{aligned}
$$

We calculate $k, \eta_{+1,0,-1}$ and $\theta_{+1,-1}$ across the PW and ZM phases by minimizing $E/N$, which also relates the condensate's momentum and longitudinal magnetization with $k = 2 F_z$. As noted for the harmonically confined quasi-1D BEC [see Fig.~\ref{phase_diagram}(a)], $k$
decreases with an increase in $\Omega$ and vanishes at the PW-ZM phase boundary. 
In this
section, we consider $c_0 n =1$ and $c_2 n = 0.1$ and quadratic Zeeman field strength $(\epsilon)=-1$ which leads
to $\Omega_{\rm c_1} \approx1.9$ and $\Omega_{\rm c_2} \approx2.75$.

\begin{figure}[!htbp]
\includegraphics[width=0.9\columnwidth]{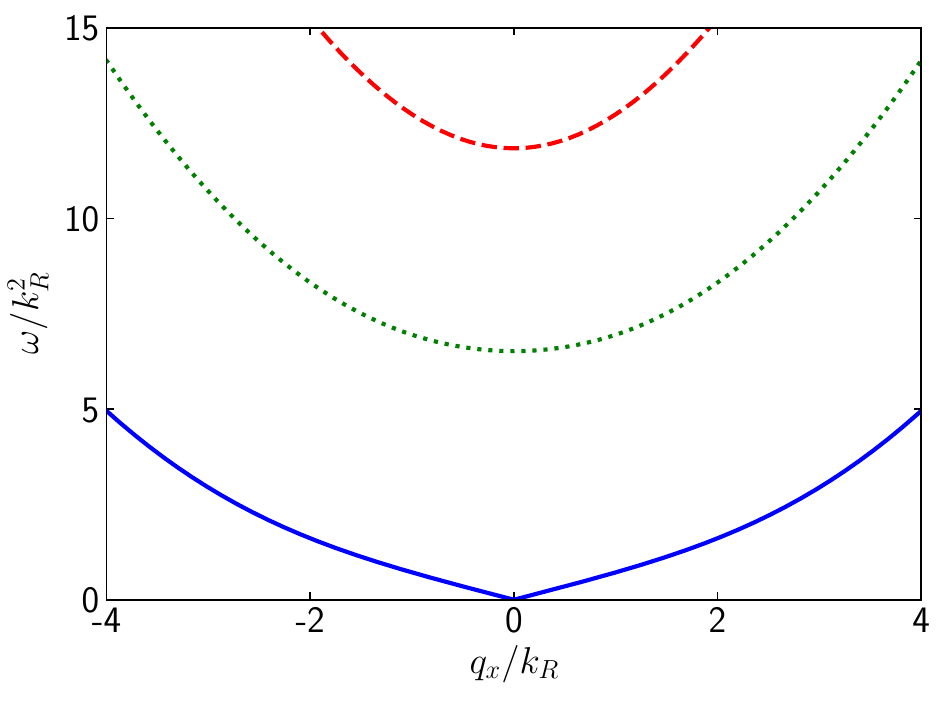}
\caption{The elementary excitation energies of a homogeneous quasi-2D SO-coupled spin-1 BEC with $c_0 n = 1$, $c_2 n = 0.1$, $\Omega=6$ and $\epsilon = -1$ in the ZM phase as a function of quasi-momentum $q_x$ for $q_{y}$ fixed to zero.}
\label{ZM_DISP}
\end{figure}

{\em Preparation of the initial state:} The PW phase breaks the ${\mathbb Z}_{2}$ symmetry with non-zero longitudinal magnetization $F_ {z}$, which 
serves as the order parameter for the phase. It is essential to prepare the initial state in the ZM phase by considering 
fluctuations to the mean-field ground state to initiate the formation of the symmetry-breaking domains after the coupling strength 
is quenched. Accordingly, we introduce noise to the ground-state solution using the truncated Wigner prescription~\cite{00018730802564254}.
To generate the noise, we consider the perturbation to the ground-state wavefunction of the ZM phase as 
$\phi_j(x,y,t) = e^{-\iota\mu t}\left[\sqrt{n_j} + \left(u^j_{\bf q} e^{\iota({\bf q}.{\bf r}-\omega t)}-v^{j*}_{\bf q}  e^{-\iota({\bf q}.{\bf r} - \omega t)}\right)/\sqrt{A}\right]$ to write the BdG equations for a homogeneous quasi-2D SO-coupled BEC
in the ZM phase, where ${\bf q} = (q_x,q_y)$ is the quasi-momentum and $u_{\bf q}$ and $v_{\bf q}$ are the quasi-particle amplitudes. We numerically solve the homogenous BdG equations to obtain the quasiparticle amplitudes and the dispersion; a typical dispersion is shown in Fig.~\ref{ZM_DISP} for $\Omega = 6$. 
The noise to be added to the mean-field ground state is 
$\delta_{j}(x,y)=\varepsilon\left(\sum_{\bf q} \left[\beta_{\bf q}^j u_{\bf q}^j e^{\iota {\bf q}. {\bf r}}-\beta_{\bf q}^{j*} v_{\bf q}^{j*} e^{-\iota {\bf q}. {\bf r}} \right]\right)$, 
where ${\bf q} = 0$ is excluded from the sum, $\beta_{\bf q}^j$ are complex Gaussian random numbers with mean and variance equal to $0$ and $1/2$, 
respectively~\cite{00018730802564254}, and $\varepsilon$ is a small real number.

\begin{figure}[!htbp]
\includegraphics[width=\columnwidth]{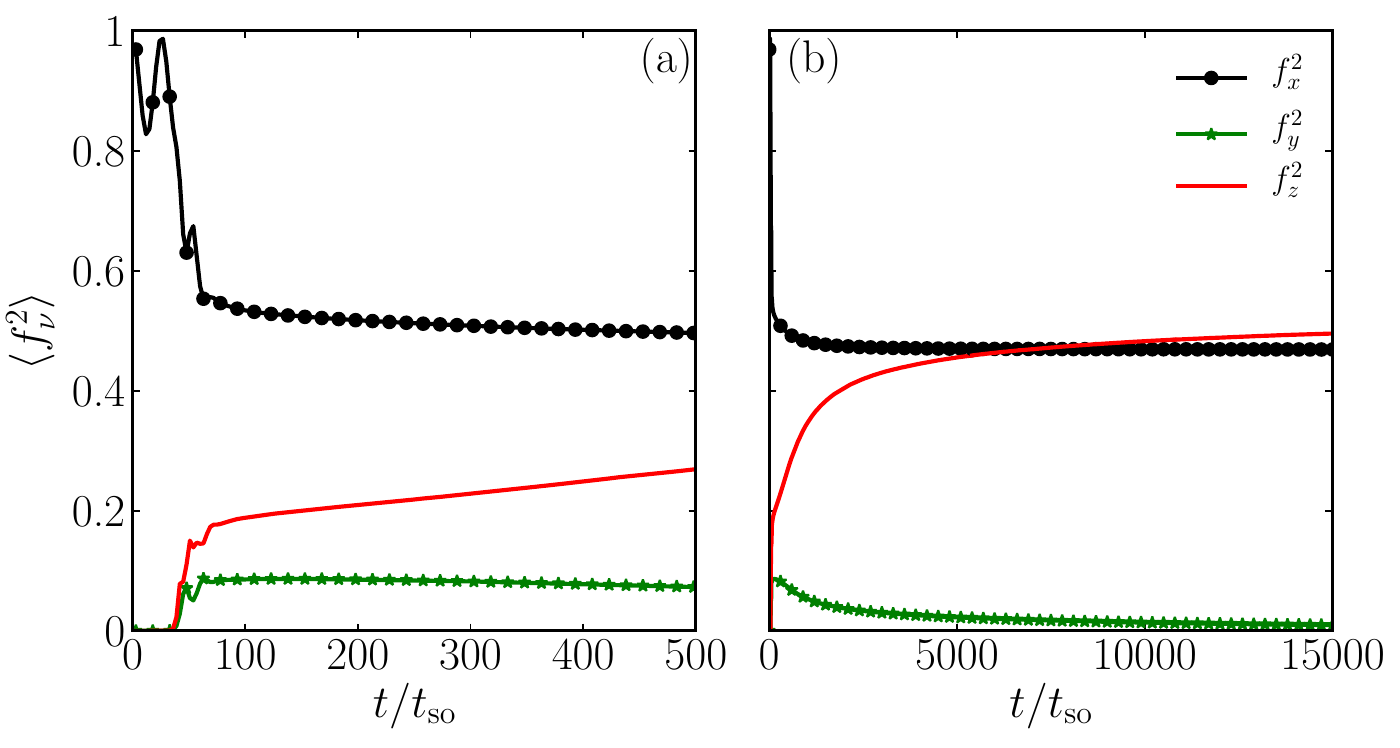}
\includegraphics[width=0.98\columnwidth]{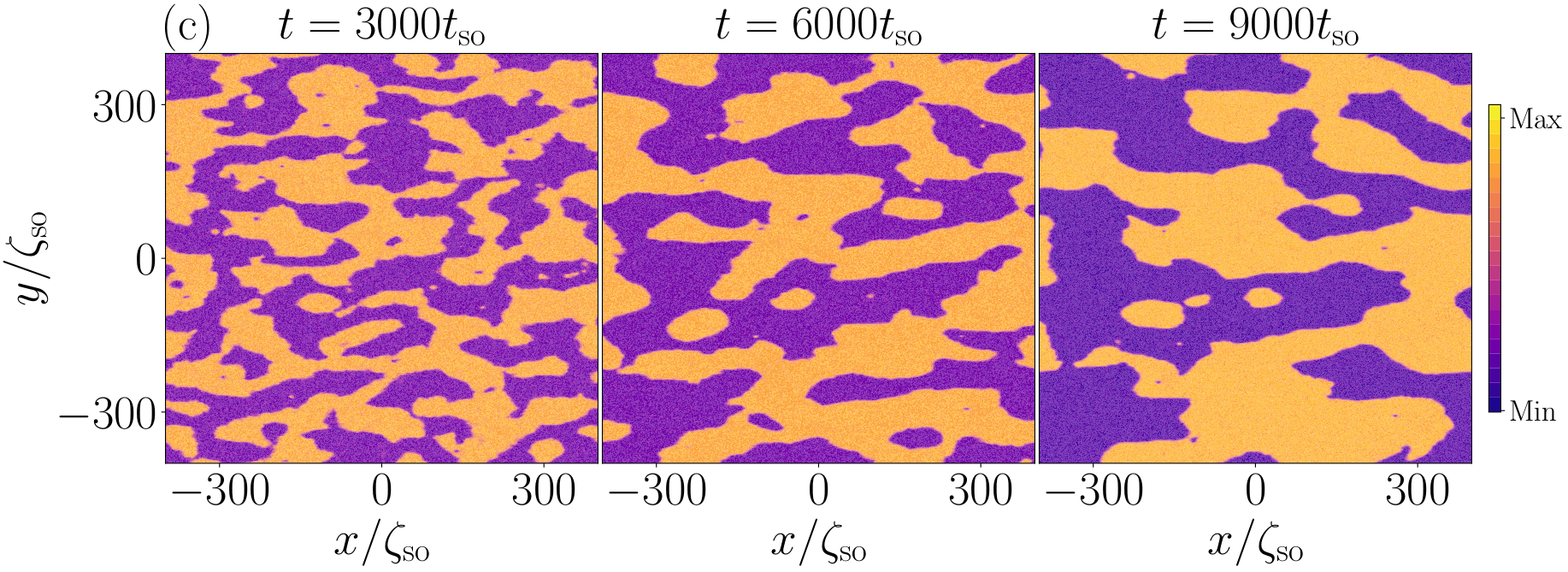}
\caption{(a) and (b): Time evolution of $\langle f_{\nu}^2\rangle$ following a quench from the ZM phase with $\Omega = 6$ to the PW phase with $\Omega = 2$. In the ZM phase, $f_x=-1$, while $f_z=0$. (c) Longitudinal magnetization density $F_z(x,y)$ or condensate momentum $k(x,y)/2$ at different times for a particular ensemble member. The domain sizes progressively increase with time.}
\label{mag}
\end{figure}

We consider a spatial 2D grid of $1024 \times 1024$ spanning a spatial extent of $800 \times 800$  with
periodic boundary conditions to simulate the quench dynamics using the GPEs for the quasi-2D system [which can obtained from Eqs. (\ref{gpe}a) and (\ref{gpe}b) by
replacing $\partial_x^2$ by $\partial_x^2+\partial_y^2$, $V(x)$ by zero and $k_{R}$ by $1$]. At $t = 0$, we suddenly quench Raman coupling to a value corresponding to which the ground-state phase is the PW phase.
After the quench, the system develops longitudinal magnetization ($f_z$). The evolution of $\langle f_{\nu}^2\rangle$ post-quench is illustrated in Fig.~\ref{mag}, where $\langle \ldots \rangle$ denotes the ensemble average with different
members of the ensemble obtained by the quench dynamics corresponding to different initial noises $\delta_j(x,y)$ added to the ground-state 
solution at $t = 0$.
We consider an ensemble average over 30 simulations. Initially, at $t = 0$, both $\langle f_y^2\rangle$ and $\langle f_z^2\rangle$ are
zero, while $\langle f_x^2 \rangle$ is equal to 1 [see Fig.~\ref{mag}(a)].
Between $t \approx 20t_{\rm so}$ to $t \approx 60 t_{\rm so}$, $\langle f_z^2\rangle$ grows exponentially; as more time elapses, the magnetization
starts to saturate steadily [see Fig.~\ref{mag}(b)], and the domain begins to coarsen; domains of the $F_z(x,y)$  or condensate momentum $k(x,y)/2$ at different times
for a single noise realization are shown in Fig.~\ref{mag}(c). The magnitude of $f_z$ at $t = 15000$ is $0.71$ and
is close to the equilibrium ground state value of $0.73$.
As the domains coarsen, the coarsening dynamics become universal and independent of the microscopic details.
By using the translational invariance of the system and isotropy of the 2D space, these domains are described by the order parameter autocorrelation function~\cite{PhysRevA.101.023608}
\begin{eqnarray}
   G(r,t) &=&   \left \langle \frac{1}{A} \left \langle \int d{\bf r}' F_z({\bf r}',t) F_z({\bf r}'+{\bf r},t)  \right\rangle_{\rm ang} \right\rangle,\nonumber\\
                &=&  \left \langle\sum_{\bf k} |\tilde{F}_z({\bf k},t)|^2 J_0(kr) \right\rangle,
\end{eqnarray}
where $\tilde{F}_z({\bf k},t)$ denotes the discrete Fourier transform of $F_z({\bf r},t)$, $J_0(kr)$ is the zeroth order Bessel function of first kind, $\langle \rangle_{\rm ang}$ denotes the angular average.
\begin{figure}[!htbp]
\includegraphics[width=0.9\columnwidth]{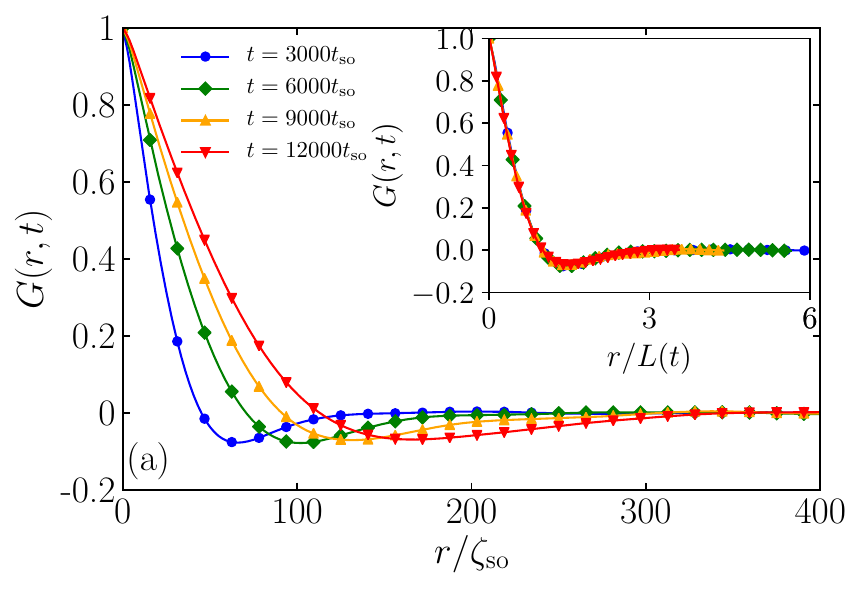}
\includegraphics[width=0.87\columnwidth]{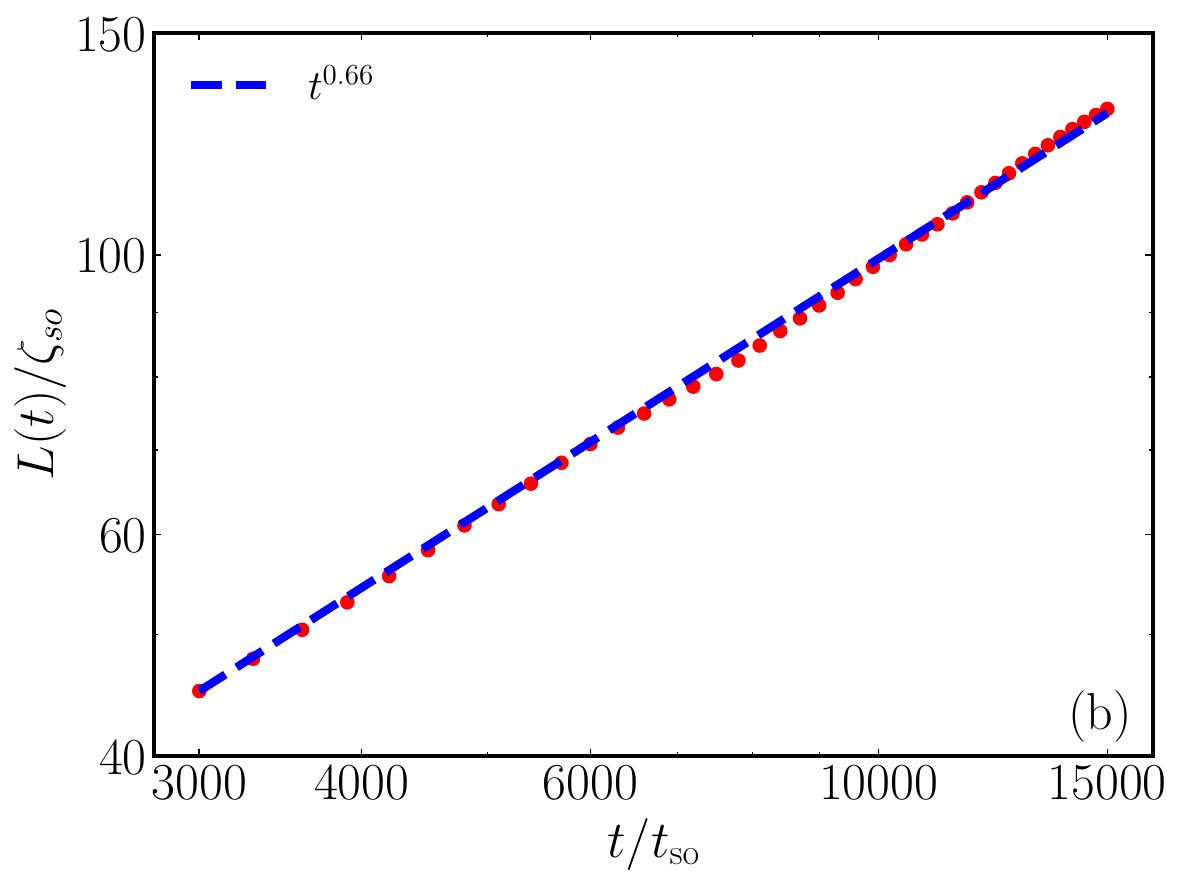}
\caption{(a) Correlation function $G(r,t)$ at different times; inset: the correlation functions $G(r/L(t),t)$ with the spatial coordinates rescaled by $L(t)$ collapsing onto a single function demonstrating the universal coarsening behavior. (b) the characteristic length $L(t)$ as a function of $t$ (red dots) and the best-fit $L(t)\sim t^{0.66}$ (dashed-blue line).}
\label{universal_combine}
\end{figure}
The correlation functions at different times are shown in Fig.~\ref{universal_combine}, and
as time elapses during the coarsening dynamics, the correlations extend over a broader spatial domain.
We define the average domain size $L(t)$ as the first zero of the correlation function $ G(r, t)$. 
When one measures the spatial coordinates in the units of $L(t)$, the correlation functions at different times fall on 
a single curve $f(r) = G(r/L(t), t)$. This demonstrates the universal coarsening behavior (see the inset of Fig.~\ref{universal_combine}).
The characteristic length $L(t)$ grows as a power law with time, as shown in Fig.~\ref{universal_combine}(b). 
We find that for a range of $\Omega$ values (2, 2.1, 2.2, and 2.3), $L(t) = (t/t_0)^{\beta} L_0$ with a dynamic critical exponent 
$\beta = 0.66$, where $L_0$ is the characteristic length scale at an arbitrary reference time $t_0$.
We find that the quenching from the ZM to the PW phase by a sudden decrease in the quadratic Zeeman 
field $\epsilon$ below $\epsilon_{\rm c_2}$ demonstrates the same power-law scaling with $\beta = 0.66$ (results not shown here).  
These results indicate that the late-time post-quench dynamics from the PW to the ZM phase in an
SO-coupled BEC belongs to a binary fluid universality class in the inertial hydrodynamic regime~\cite{PhysRevA.31.1103,huh2024universality, PhysRevA.88.013630, PhysRevLett.113.095702, PhysRevLett.116.025301}.

\section{Summary and Conclusions}
\label{Sec-V}
In the first part of this study, we examined the collective excitations of a harmonically trapped quasi-1D SO-coupled
spin-1 BEC. We analyzed the dependence of these excitations on two experimentally controllable parameters: the Raman coupling and the Zeeman field strengths.
By examining the behavior of excitation modes, we identified key signatures of phase transitions between different quantum phases.
We calculated dipole and breathing modes for density and spin channels by applying appropriate perturbations to the system. The spin-dipole 
and spin-breathing modes soften near the ST1-PW phase boundary, whereas the density-dipole and density-breathing modes soften near 
the PW-ZM phase. At lower values of Raman coupling strength, the system can show a direct transition from the ZM to the ST1 phase with a variation in
the quadratic Zeeman field. In this case, as the 
system approaches the transition to the ST1 phase, we observe the softening of a symmetric double roton mode, indicating the system's
tendency towards crystallization. Furthermore, we confirmed the order of the phase transitions by examining momentum and the spin expectation 
per particle of the condensate across the three phases, which agree with the behavior of the collective
excitations, especially the two density modes, across the transition points.

In the second part of this study, we focused on the universal coarsening dynamics of a homogeneous quasi-2D SO-coupled
spin-1 BEC by quenching Raman coupling (or quadratic Zeeman field strength) from the ZM to the PW phase. We demonstrated
that the correlation function of the order parameter displays dynamic scaling during the late-time dynamics,
allowing us to determine the dynamic critical exponent.
We showed the formation of magnetic domains and the universal behavior of the correlation function of order parameters as it
scales by a characteristic length $L(t)$. 
This characteristic length increases with time following a power law $ L(t) \sim t^{0.66}$. This finding is consistent with the inertial hydrodynamics domain growth
law of binary fluids.

\subsection*{Acknowledgements}
We acknowledge the National Supercomputing Mission (NSM) for providing the computing resources
of 'PARAM SMRITI' at NABI, Mohali, and 'PARAM Ananta' at IIT Gandhinagar, which is implemented by C-DAC and supported by the Ministry of
Electronics and Information Technology (MeitY) and Department of Science and Technology (DST), Government
of India. SG acknowledges support from the Science and Engineering Research Board, Department
of Science and Technology, Government of India through Project No. CRG/2021/002597.

 \bibliography{bib_file.bib}{}
 \bibliographystyle{apsrev4-1}
\end{document}